# SDR Driver for Precise Timing Applications

**Fabrizio Pollastri** F.POLLASTRI@INRIM.IT
INRIM, strada delle cacce 91, 10135 Torino, Italy

## Abstract

This paper describes the development of software drivers that enable precise timing applications with the low-cost HackRF One software-defined radio (SDR). The original firmware was extended with timing functions for synchronized sampling, time-scale control, and clock-frequency adjustment. An initial implementation based on a Python wrapper was later redesigned as a SoapySDR-compatible driver. Using an AI-assisted development workflow, the new driver was generated from the existing HackRF One and SoapySDR codebases and became operational after only limited debugging. The approach reduced development effort by more than an order of magnitude while improving software maintainability.

## 1. Introduction

The main applications of SDRs (Software Defined Radios) involve the transmission of data, voice, and images. A less common SDR application is precise timing, which is used to transfer time between different locations with high accuracy.
When precision timing comes into play, one immediately thinks of using high-end SDRs that allow synchronization with a reference frequency and PPS (Pulse Per Second). However, even a low-cost SDR can be used for accurate synchronization systems. The HackRF One (HackRF One, 2026), a low-cost SDR with open source firmware and open hardware documentation, was used to build an accurate time transfer system (Pollastri, 2024) by providing the SDR with a modified firmware. The original firmware was extended with commands for timed and synchronized signal sampling. This extended firmware also requires an extended driver to exploit the new SDR commands.
This work focuses on the initial manual development of the driver and its subsequent evolution into a SoapySDR-compatible (SoapySDR, 2026) driver developed with AI assistance (SoapyHackRFHTime, 2026).

### 2.1 SDR timing requirements

When an SDR is used for timing applications, the main requirement is to have each signal sample traceable to a given time scale. To obtain this, the SDR must have:

- a time scale with seconds and ticks counters
- a sampling clock synchronous with the time scale
- a fine adjustment of the time scale to track a reference signal.

All of these features were absent from standard HackRF One firmware. Fortunately, the HackRF One has enough unused onboard hardware resources that can be activated by an extended firmware (hackrf-htime, 2023-2026) to fulfill all these requirements.

### 2.2. SDR driver extensions

The API of the original driver was extended to provide access to the requested timing functions as described below. Figure 1 shows a functional diagram of the SDR and the hardware/firmware blocks controlled by the API extensions.

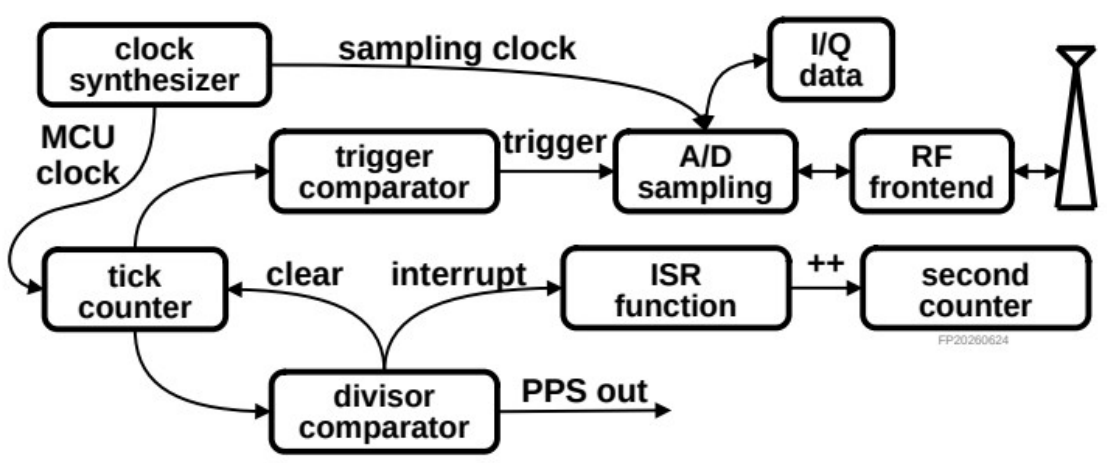

*Figure 1: functional diagram of extended API.*

The time scale is implemented by the **tick counter**, which runs at the MCU clock rate, and by the **second counter**, which is incremented every second by the **ISR function**, as shown in Figure 1. The time scale second and tick values are set and read by the functions:

- **get_seconds_now**
  get current seconds counter value immediately;
- **set_seconds_now**
  set second counter value immediately;
- **get_ticks_now**
  get ticks counter value immediately;
- **set_ticks_now**
  set ticks counter value immediately.

These functions can be called to initialize the time scale and to read the current time of the SDR during operation.
When the seconds value needs to be changed with a known phase with respect to the PPS, the following function is used

- **set_seconds_next_pps**
  set seconds counter value at next PPS.

To configure the SDR for sampling the signal synchronously with the time scale there is the following function:

- **set_mcu_clk_sync**
  synchronize the SDR sampling and MCU clocks to SDR main clock.

This function reconfigures the **clock synthesizer** to provide the same clock source to the signal-sampling A/D and to the SDR MCU, where the time scale is implemented by the **tick counter**. This avoids phase slippage between the MCU counters/timers and the sampling clock, as well as phase uncertainties and shifts between the sampled data and the SDR time scale.
The one-second period of the time scale is generated by dividing the 200 MHz MCU clock using the **divisor comparator** shown in Figure 1. The divisor is set by the function

- **set_divisor_next_pps**
  Set the tick counter divisor at the next PPS.

To obtain a 1 s period, the divisor is set to $2 \cdot 10^8$-1.
The timing application that uses the extended functionalities of the SDR was designed for time transfer between a reference SDR receiver synchronized to a reference frequency and PPS, for example from an atomic clock, and another SDR, possibly remote. This implies the capability for the remote SDR to continuously adjust its internal clock to remain locked to the reference frequency. This capability is given by the function

- **set_clk_freq**
  fine adjustment of SDR main clock frequency.

This function configures the frequency parameters of the **clock synthesizer**. Normally, this function provides fine clock-frequency adjustments in the range ±100 Hz at 10 MHz sampling rate.
At system startup or after a signal interruption, the SDR timescale may differ significantly from the reference timescale then fine frequency adjustments are insufficient. So, the extended API provides a function to implement a coarse phase jump of the SDR timescale.

- **set_divisor_one_pps**
  Set the tick counter divisor from next PPS for one counter cycle then restore previous divisor.

For a single PPS cycle, the **divisor comparator** changes the PPS period to produce a specified phase jump in the SDR time scale. The jump amplitude is computed from the phase difference between the local timescale and the remote reference timescale.
The timing application requires a burst of received-signal samples every second with a known phase relationship to the SDR time scale. The sampling start time is therefore set by the **trigger comparator** through the following function:

- **set_trig_delay_next_pps**
  Set sampling trigger delay from next start of second.

Together, these functions provide the synchronization, time-scale control, frequency adjustment, and timed-sampling capabilities required to use the HackRF One for precision timing applications. More details about the API extensions can be found in the extended firmware documentation (hackrf-htime, 2023-2026).

### 2.3. Python wrapper

The original SDR driver was written in C, and the extensions were therefore implemented in the same language. Since the application software was written in Python, an intermediate interface layer between Python and C was required. In this first driver implementation, an existing Python wrapper for the HackRF One was chosen (pyhackrf2, 2023). Like the driver, the wrapper was extended to manage the additional functions.

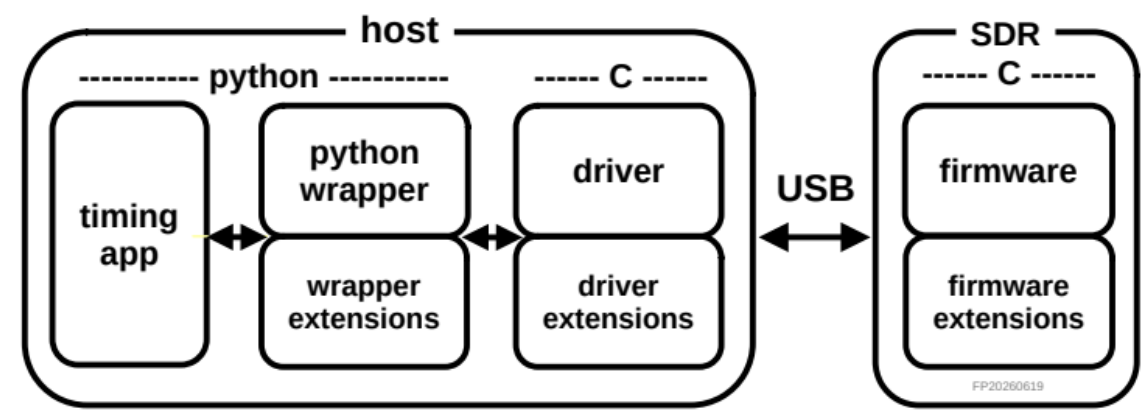

*Figure 2: main components of the timing application.*

Figure 2 shows the architecture of the main software components used in the system of the timing application (Pollastri, 2024). During the driver development, the presence of a Python wrapper made the work harder, because it had to be updated whenever the API changed in the firmware and in the driver.

### 3.1 New version of SDR driver

The availability of AI programming agents provided a strong incentive for a complete rewrite of the driver.
One of the main goals of the new driver version was to eliminate the need for a Python wrapper. A suitable solution for this purpose is offered by the Soapy SDR library

(SoapySDR, 2026).
As stated in the documentation, “SoapySDR is an open-source generalized API and runtime library for interfacing with SDR devices.” Among many interesting features, it provides Python bindings for the driver API without any additional effort.

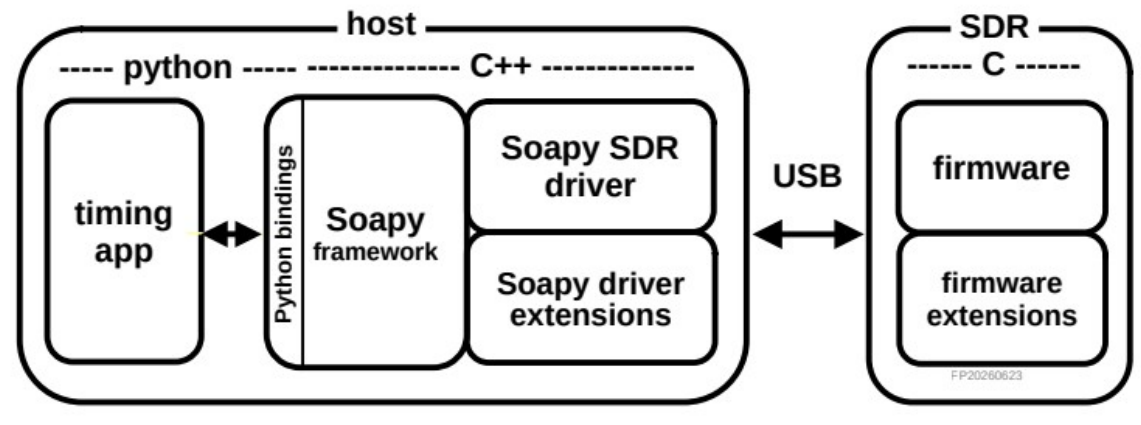

*Figure 3: main components of timing application after driver rewrite.*

It supports several drivers, written in C++, for most commercially available SDRs. The HackRF One SDR is also supported by the driver SoapyHackRF (SoapyHackRF, 2026). To benefit from the features of the SoapySDR framework, a complete rewrite of the first version of the HackRF One driver was needed.

### 3.2 AI environment setup

Since all the code involved in this driver development is based on GitHub repositories, an easy choice was to use the Copilot AI agent (GitHub Copilot, 2026) integrated with Visual Studio IDE. As a first step, the IDE was provided with a copy of each repo folder relevant in the development. To achieve this goal, an inventory of the relevant repositories was compiled, resulting in the following set of software assets.

- **SoapySDR,**
  the framework library;
- **SoapyHackRF,**
  the original driver from the Soapy SDR *framework* for the HackRF SDR;
- **hackrf/host,**
  the original driver version of the HackRF One firmware;
- **hackrf-htime/host,**
  the first driver version with the timing extensions of the HackRF One firmware.

### 3.2 AI prompt strategy

Since an AI agent was used to facilitate development of the new driver, the next step was to write an AI prompt. An effort was made to find a better balance between detailed specifications and a high level description.

3.2.1. Timing extensions

The question for the first part of the AI prompt was how to give the detailed specifications of the timing extensions at the highest possible level. The availability of the repositories of both the original HackRF One driver and the extended one, allowed the following statement to be included in the prompt: in hackrf-htime/host there is an extended version of the driver in hackrf/host for the SDR HackRF One. This is sufficient for the AI agent to infer all implementation details of the extensions from the code differences between the original driver and the extended one.

3.2.2. Soapy implementation

The second part of the prompt describes how the extensions should be implemented as a Soapy driver. Since the SoapySDR framework already includes a driver for the HackRF One (SoapyHackRF), this is the natural starting point to which the timing extensions could be added. This part of the prompt can be written as: create a new driver by adding the above time extensions to the driver SoapyHackRF. Two further constraints were added to facilitate future maintenance: put the code of the timing extensions and the original code in different files and different classes. The classes are structured as base classes with the original driver code and child classes with the timing extensions code. Table 1 shows this scheme.

Table 1: Files and classes in the new extended driver

| Original code | Extensions code |
|---|---|
| **Files** | **Files** |
| | HackRFHTime_Registration.cpp |
| HackRF_Session.cpp | HackRFHTime_Session.cpp |
| HackRF_Settings.cpp | HackRFHTime_Settings.cpp |
| HackRF_Streaming.cpp | HackRFHTime_Streaming.cpp |
| SoapyHackRF.hpp | SoapyHackRFHTime.hpp |
| **Base Classes** | **Child Classes** |
| SoapyHackRFSession | SoapyHackRFHTimeSession |
| SoapyHackRF | SoapyHackRFHTime |

Finally, to help the AI agent to have the proper context, the SoapySDR framework files were also provided.

### 3.3. Generated Driver

The AI agent created the new SoapyHackRFHTime module by copying the original SoapyHackRF code, refactoring the driver identity and build target, and adding HTime time/setting support while keeping the original settings/streaming APIs intact. Updated documentation, packaging metadata, versioning, and tests to match the new hackrf-htime driver id and HTime API.

3.3.1. Debug operations

The debugging phase required after the initial AI-

generated implementation was limited and consisted mainly of integrating the driver correctly into the SoapySDR environment rather than redesigning its architecture. The first execution of the test suite revealed a device instantiation failure caused by an incorrect device selection strategy. After adapting the test to follow the enumeration procedure used by the existing SoapySDR examples, the driver was correctly detected, but execution exposed a series of segmentation faults. These problems were resolved through a small number of iterative debugging cycles that included removal of conflicting development libraries, reordering of streaming activation and deactivation calls, restoration of hardware synchronization support, alignment with the current libhackrf version (libhackrf, 2026), and implementation of a missing clock-output configuration setting. The development log reported by the AI agent shows that the generated fixes were focused on isolated integration issues rather than substantial modifications of the generated architecture. This indicates that the AI-generated driver was structurally consistent with the intended architecture. The remaining work consisted mainly of resolving runtime compatibility and integration issues before the driver passed the functional tests.

### 3.4. Driver Comparison

Table 2 summarizes the main differences between the traditional development approach and the AI-assisted development workflow adopted in this work.

*Table 2:* Comparison between the Python-wrapper and SoapySDR drivers

| **Feature** | **Python Wrapper Driver** | **SoapySDR Driver** |
|---|---|---|
| Language | C and Python | C++ |
| Python bindings | Manual | Automatic |
| Maintenance effort | High | Low |
| AI-generated | No | Yes |
| Soapy compatibility | No | Yes |

The comparison shows that the principal advantage of AI is not only the reduction in implementation time, but also the ability to reuse and integrate existing software components with limited knowledge of the underlying framework. Most of the developer effort shifted from writing implementation code to defining the prompt, validating the generated solution, and performing a limited number of debugging iterations. The generated driver preserved a modular architecture compatible with the SoapySDR framework and required only minor corrections to resolve integration issues. These observations suggest that, for software projects based on mature open-source codebases, AI agents can substantially accelerate development while maintaining software quality and long-term maintainability.

### 3.5. Conclusions

This work presented the evolution of a custom HackRF One driver with timing extensions from a traditional C/Python implementation to a SoapySDR-compatible driver developed with the assistance of an AI agent. The resulting driver eliminated the need for a dedicated Python wrapper while preserving all timing functions required for precision time-transfer applications. After only a limited number of debugging iterations and minor refactoring, the generated software became fully operational, reducing the overall development effort by more than an order of magnitude.

An important lesson learned is that AI agents are particularly effective when a project is based on existing, well-structured open-source software. In this work, the availability of the original HackRF One driver, the extended timing driver, and the SoapyHackRF implementation enabled the AI system to infer most implementation details directly from the existing code, allowing the developer to focus on architectural decisions, prompt design, and validation rather than framework-specific programming.

A second lesson learned concerns aspects of prompt engineering that are easily overlooked because they lie outside the source code itself. The first is the importance of explicitly considering the host environment in which the generated software will be built and tested. An AI agent can correctly generate the implementation while making implicit assumptions about library versions, installation paths, build configuration, or runtime settings that differ from the actual development environment, resulting in integration failures that are not caused by the generated driver itself.

A third lesson, also easily overlooked, is the need to keep the development environment clean throughout successive AI-assisted iterations. During the debugging process, obsolete libraries, previous builds, or partially installed driver versions accumulated in the system and occasionally interfered with testing, producing misleading failures unrelated to the current implementation. Explicitly instructing the AI agent to verify the execution environment and periodically remove obsolete development artifacts proved to be as important as specifying the software requirements themselves.

At the same time, the experience confirms that AI assistance does not eliminate the need for human supervision. Careful prompt engineering, review of the generated driver, debugging of integration issues, and final validation remained essential steps to obtain a reliable driver. The approach therefore shifts the developer's effort from code implementation toward specification and verification.

The results indicate that AI-assisted software development can substantially accelerate the migration and extension of existing software projects while preserving maintainability and modularity. Similar workflows may therefore be applicable to other SDR drivers and to software projects that involve integrating existing codebases into established software frameworks.

∎